\documentclass[11pt,twocolumn]{article}
\usepackage{appendix}
\usepackage{amsmath}
\usepackage{amssymb}
\usepackage{amsfonts}
\usepackage{amscd}
\usepackage{delarray}
\usepackage[font=footnotesize,format=plain,labelfont=bf,up,]{caption}
\usepackage{sidecap}
\usepackage{graphicx}
\usepackage[T1]{fontenc}
\usepackage[utf8]{inputenc}
\usepackage[francais,british]{babel}
\usepackage{color}
\usepackage{tikz}
\usepackage{bbm}
\usepackage{authblk}
\usepackage{enumitem}
\usepackage{abstract}
\usepackage{hyperref}
\usepackage{subcaption}

\makeatletter

\definecolor{blue}{rgb}{0,0,.8}
\definecolor{black}{rgb}{0,0,0}
\definecolor{purplerep}{rgb}{1,0.1,1}

\newcommand{\micron}{\,\mu\textrm{m}}
\newcommand{\mm}{\,\textrm{mm}}
\newcommand{\nm}{\,\textrm{nm}}
\newcommand{\cm}{\,\textrm{cm}}
\newcommand{\GHz}{\,\textrm{GHz}}
\newcommand{\MHz}{\,\textrm{MHz}}
\newcommand{\ms}{\,\textrm{ms}}

\newcommand{\pOhm}{\,\textrm{p}\Omega}
\newcommand{\muOhm}{\,\mu\Omega}

\newcommand{\Rmnum}[1]{\expandafter\@slowromancap\romannumeral #1@}

\makeatother

\def\idmx{1\!{\rm l}}

\let\oldsqrt\sqrt
\def\sqrt{\mathpalette\DHLhksqrt}
\def\DHLhksqrt#1#2{
\setbox0=\hbox{$#1\oldsqrt{#2\,}$}\dimen0=\ht0
\advance\dimen0-0.2\ht0
\setbox2=\hbox{\vrule height\ht0 depth -\dimen0}
{\box0\lower0.4pt\box2}}
\newcommand{\tens}[1]{\smash{\underline{\underline{#1}}}}
\newcommand{\re}{\mathfrak{Re}}

\begin{document}

\title{Absorption in quantum electrodynamics cavities\\in terms of a quantum jump operator}
\date{\today}

\author{V. Debierre}
\author{G. Demésy}
\author{T. Durt}
\author{A. Nicolet}
\author{B. Vial}
\author{F. Zolla}

\affil{Aix Marseille Université, CNRS, École Centrale de Marseille, Institut Fresnel, UMR 7249, 13013 Marseille France.}

\twocolumn[
\maketitle
\vglue -1.8truecm
\vspace{20pt}
\begin{abstract}
We describe the absorption by the walls of a quantum electrodynamical cavity as a process during which the elementary excitations (photons) of an internal mode of the cavity exit by tunneling through the cavity walls. We estimate by classical methods the survival time of a photon inside the cavity and the quality factor of its mirrors.
\end{abstract}
\vspace{20pt}
]

\section{Introduction} \label{sec:Intro}

The absorption of a photon by the walls and/or the escape of the photons from a cavity is an intrinsically irreversible and non-unitary process. Its deep understanding would in principle require an accurate description of the properties of the environment (geometry and quality of the mirrors in the cavity, possible escape towards the empty space outside the cavity and so on). However this description is too complicated to be performed in practice and, commonly, one derives the effective master equation that incorporates the losses inside the cavity thanks to an ``educated guess''. In this \emph{ad hoc} description only one free parameter, the survival time of a single photon inside the cavity, characterizes the losses. The value of this parameter is fixed in last resort by experiments \cite{HarocheRaimond}.

The aim of our work is to refine the usual description of a lossy cavity by 
\begin{enumerate}
\item{describing the absorption and/or escape by the boundaries of a QED cavity as a tunneling process during which elementary excitations (photons) of an internal mode of the cavity exit the cavity by tunneling and}
\item{estimating by classical methods the survival time of a photon inside the cavity.}
\end{enumerate}

The paper is organized as follows.

In sect.~\ref{sec:Jump}, we introduce a Quantum Jump Operator aimed at simulating the escape of the photons leaving the cavity. We show in the framework of this model that the survival time of a photon inside the cavity is in last resort defined by a single free parameter, which measures the coupling between a photon inside the cavity and a photon having the same frequency and leaving the cavity by tunneling. This result is derived, in the weak coupling regime, along the lines of the so-called Friedrichs model \cite{Friedrichs} which is very close in mind to the Wigner-Weisskopf treatment of exponential decay. 
 
An interesting by-product of our approach (sect.~\ref{sec:Coherent}) is that it allows us to show that coherent states of the cavity couple to the environment without getting entangled to it. Usually this property is derived by solving the master equation that describes the losses in the cavity. In our case this property is directly linked to the nature of the coupling between in and out modes as expressed by our quantum jump operator.
 
Since, on the other side, coherent states are quasi-classical in the sense that they minimize Heisenberg-type uncertainty relations, it is consistent to neglect the quantum fluctuations in order to obtain a first estimation of the single free parameter that is not addressed by our quantum jump model, essentially the absorption rate of the cavity. In this coarse grained approach where the mode inside the cavity is treated as a classical oscillator, it is justified to resort to the classical Maxwell's equations in order to estimate the losses. By doing so we attempt to estimate (sect.~\ref{sec:Maxwell}) the lifetime of a photon inside the cavity, and obtain an indirect estimation of the ``effective'' quality factor of the mirrors used in the QED cavity \cite{Pipeau} developed by the group of Serge Haroche at École Normale Supérieure in Paris \cite{HarocheRaimond,CavityQED,Pipeau,HarocheNature}. Our conclusion (sect.~\ref{sec:Ccl}) is that the losses are essentially due to the geometry of the cavity, and not to the absorption of the mirrors.

\section{Tunneling processes in terms of a quantum jump operator} \label{sec:Jump}

In this section we introduce a simple, one-dimensional, model aimed at simulating the losses of a quantum electrodynamics cavity. We study its along the lines of the Wigner-Weisskopf time-dependent perturbation theory.

\subsection{A quantum jump operator} \label{sec:QJO}

Even for a cavity with perfectly conducting mirrors, there would be losses due to the open geometry such as that of the cavity used by Haroche and coworkers \cite{HarocheRaimond,HarocheNature}, which explains the finite lifetime of the modes. In this section we make the approximation that the only relevant cavity mode is the fundamental mode. We take the cavity to be in first approximation closed, so that the mode we will deal with will be the fundamental mode (with frequency $\omega_0$) of what is the electrodynamical equivalent of an infinite potential well, but we then weakly couple this mode to the continuous infinity of modes that lie outside the cavity. This allows us to account for the losses in a simplified way. In the Schrödinger picture our Hamiltonian reads (everywhere integrals of the type $\int\mathrm{d}k\,f\left(k\right)$ are to be understood as integrals over the whole real $k$-line)
\begin{align} \label{eq:hamham}
\hat{H}_{\mathrm{S}}&=\hbar\omega_0\,\hat{a}_{\mathrm{S}}^\dagger\,\hat{a}_{\mathrm{S}}\otimes\hat{\idmx}_{\mathrm{out}}+\hat{\idmx}_{\mathrm{in}}\otimes\int\mathrm{d}k\left|k\right|\hbar c\,\hat{b}_{\mathrm{S}}^\dagger\left(k\right)\hat{b}_{\mathrm{S}}\left(k\right)\nonumber\\
&+\!\int\mathrm{d}k\left[\lambda\left(\left|k\right|\right)\hat{a}_{\mathrm{S}}\,\hat{b}_{\mathrm{S}}^\dagger\left(k\right)+\lambda^*\left(\left|k\right|\right)\hat{a}_{\mathrm{S}}^\dagger\,\hat{b}_{\mathrm{S}}\left(k\right)\right]
\end{align}
where the (Schrödinger picture) ladder operators for the cavity ($\hat{a}_{\mathrm{S}}$ and $\hat{a}_{\mathrm{S}}^\dagger$) and external world ($\hat{b}_{\mathrm{S}}\left(k\right)$ and $\hat{b}_{\mathrm{S}}^\dagger\left(k\right)$) obey the following commutation relations:
\begin{subequations}
\begin{align}
\left[\hat{a}_{\mathrm{S}},\hat{a}_{\mathrm{S}}^\dagger\right]&=\hat{\idmx},\\
\left[\hat{b}_{\mathrm{S}}\left(k\right),\hat{b}_{\mathrm{S}}^\dagger\left(q\right)\right]&=\delta\left(k-q\right)\,\hat{\idmx}
\end{align}
\end{subequations}
with all other commutators being zero. The second integral on the right-hand-side of (\ref{eq:hamham}) is what we called a jump operator. Despite this terminology it should be clear that the evolution is unitary. The so-called jump operator simply describes the escape of the photons outside the cavity, no actual (non-unitary) jumps are involved here, in contrast to the usual master equation treatment. The positive (negative) wavelength modes propagate in the right (left) outside region and are associated to plane waves $\idmx_{\mathrm{rg}}\,\mathrm{e}^{-\mathrm{i}kx}$ ($\idmx_{\mathrm{lf}}\,\mathrm{e}^{+\mathrm{i}kx}$), where $\idmx_{\mathrm{rg}}$ and $\idmx_{\mathrm{lf}}$ are indicator functions for the right and left regions respectively. Introducing the size $L$ of the cavity, they are simply given by
\begin{equation} \label{eq:Oliver}
\idmx_{\mathrm{rg/lf}}\left(x\right)=\Theta\left(\pm x-\frac{L}{2}\right).
\end{equation}

\subsection{Heisenberg equations for the ladder operators} \label{sec:Werner}

We now switch to the Heisenberg picture. Since the Hamiltonian in the Schrödinger picture has no explicit time dependence, the evolution operator has the standard form $\hat{U}\left(t\right)=\exp\left(-\frac{\mathrm{i}}{\hbar}\hat{H}_{\mathrm{S}}t\right)$. The operators in the Heisenberg picture are thus defined by $\hat{A}_{\mathrm{H}}\left(t\right)\equiv\hat{U}^\dagger\left(t\right)\hat{A}_{\mathrm{S}}\left(t\right)\hat{U}\left(t\right)$. They obey the Heisenberg equations
\begin{subequations}
\begin{equation} \label{eq:Heisenberg}
\frac{\mathrm{d}\hat{A}_{\mathrm{H}}}{\mathrm{d}t}=\frac{\mathrm{i}}{\hbar}\left[\hat{H}_{\mathrm{S}},\hat{A}_{\mathrm{H}}\left(t\right)\right]+\hat{U}^\dagger\left(t\right)\frac{\mathrm{d}\hat{A}_{\mathrm{S}}}{\mathrm{d}t}\hat{U}\left(t\right).
\end{equation}
In the case of the ladder operators which are time-independent in the Schrödinger picture, the right-hand side of (\ref{eq:Heisenberg}) reduces to the commutator:
\begin{equation} \label{eq:Liouville}
\frac{\mathrm{d}\hat{A}_{\mathrm{H}}}{\mathrm{d}t}=\frac{\mathrm{i}}{\hbar}\left[\hat{H}_{\mathrm{S}},\hat{A}_{\mathrm{H}}\left(t\right)\right].
\end{equation}
\end{subequations}
From the readily shown equality
\begin{equation} \label{eq:Easy}
\left[\hat{H}_{\mathrm{S}},\hat{A}_{\mathrm{H}}\left(t\right)\right]=\hat{U}^\dagger\left(t\right)\left[\hat{H}_{\mathrm{S}},\hat{A}_{\mathrm{S}}\left(t\right)\right]\hat{U}\left(t\right)
\end{equation}
one gets the following Heisenberg equations for the ladder operators:
\begin{subequations} \label{eq:Evolution}
\begin{align}
\frac{\mathrm{d}\hat{a}_{\mathrm{H}}}{\mathrm{d}t}&=-\mathrm{i}\omega_0\,\hat{a}_{\mathrm{H}}\left(t\right)-\frac{\mathrm{i}}{\hbar}\int\mathrm{d}k\,\lambda^*\left(\left|k\right|\right)\hat{b}_{\mathrm{H}}\left(k,t\right), \label{eq:Evola}\\
\frac{\partial\hat{b}_{\mathrm{H}}}{\partial t}\left(k,t\right)&=-\frac{\mathrm{i}}{\hbar}\lambda\left(\left|k\right|\right)\hat{a}_{\mathrm{H}}\left(t\right)-\mathrm{i}c\left|k\right|\hat{b}_{\mathrm{H}}\left(k,t\right)
\end{align}
\end{subequations}
and the adjoint evolution equations for the respective adjoint operators.

\subsection{Time-dependent perturbation theory in the weak coupling regime} \label{sec:WigWei}

Forcing an ansatz expansion of $\hat{a}_{\mathrm{H}}\left(t\right)$ in terms of Schrödinger picture-destruction operators of the type
\begin{equation} \label{eq:Ansatz}
\hat{a}_{\mathrm{H}}\left(t\right)=\alpha\left(t\right)\hat{a}_{\mathrm{S}}+\int\mathrm{d}k\,\beta\left(k,t\right)\hat{b}_{\mathrm{S}}\left(k\right),
\end{equation}
which is easily seen to be compatible with (\ref{eq:Evolution}), we then get the following system of coupled differential equations:
\begin{subequations} \label{eq:Aristocats}
\begin{align}
\frac{\mathrm{d}\alpha}{\mathrm{d}t}&=-\mathrm{i}\omega_0\alpha\left(t\right)-\frac{\mathrm{i}}{\hbar}\int\mathrm{d}k\,c\,\lambda^*\left(\left|k\right|\right)\beta\left(k,t\right), \label{eq:Fora}\\
\frac{\partial\beta}{\partial t}\left(k,t\right)&=-\frac{\mathrm{i}}{\hbar}\lambda\left(\left|k\right|\right)\alpha\left(t\right)-\mathrm{i}c\left|k\right|\beta\left(k,t\right). \label{eq:Forb}
\end{align}
\end{subequations}
As is clear from $\hat{a}_{\mathrm{H}}\left(t\right)=\hat{U}^\dagger\left(t\right)\hat{a}_{\mathrm{S}}\left(t\right)\hat{U}\left(t\right)$, the initial conditions are
\begin{equation} \label{eq:Start}
\left\{
\begin{array}{llcl}
&\alpha\left(t=0\right)&=&1,\\
\forall\,k&\beta\left(k,t=0\right)&=&0.
\end{array}
\right.
\end{equation}
When taken into account in the system (\ref{eq:Aristocats}), these initial conditions yield the following evolution equations:
\begin{subequations}
\begin{align}
\frac{\mathrm{d}}{\mathrm{d}t}\left(\mathrm{e}^{\mathrm{i}\omega_0t}\alpha\left(t\right)\right)&=-\int\mathrm{d}k\left|\frac{\lambda\left(\left|k\right|\right)}{\hbar}\right|^2\nonumber\\
&\hspace{30pt}\int_0^t\mathrm{d}t'\,\mathrm{e}^{\mathrm{i}\left(\omega_0-c\left|k\right|\right)\left(t-t'\right)}\left(\mathrm{e}^{\mathrm{i}\omega_0t'}\alpha\left(t'\right)\right),\label{betab}\\
\beta\left(k,t\right)&=-\frac{\mathrm{i}}{\hbar}\lambda\left(\left|k\right|\right)\int_0^t\mathrm{d}t'\,\mathrm{e}^{-\mathrm{i}c\left|k\right|\left(t-t'\right)}\alpha\left(t'\right)\label{betaa}. 
\end{align}
\end{subequations}
This system does not possess any known analytic solution \cite{Cohen} but it is commonly solved perturbatively, {\it \`a la} Wigner-Weisskopf  \cite{Englert} as we shall do now. The basic approximation is to consider that, since the free (non perturbed) Hamiltonian of the inside region reads $\hat{H}_{\mathrm{in}}=\hbar\omega_0\,\hat{a}_{\mathrm{S}}^\dagger\hat{a}_{\mathrm{S}}$, we can reasonably assume, in the weak coupling regime, for which $\lambda$ is a small parameter (such that the decay constant that shall arise from perturbation theory is small compared to $\omega_0$), that $\alpha$ is the product of an oscillating exponential of frequency $\omega_0$ with a function $f$ the variations of which are negligible on a time-scale of $\omega_0^{-1}$:
\begin{equation} \label{eq:Alpha}
\alpha\left(t\right)=\mathrm{e}^{-\mathrm{i}\omega_0t}f\left(t\right).
\end{equation}
The first consequence of this assumption is that energy is conserved in a first approximation, as can be seen by integrating (\ref{betaa}) considering that $\alpha\left(t\right)\simeq\mathrm{e}^{-\mathrm{i}\omega_0t}f\left(t=0\right)$ which yields
\begin{equation*}
\beta\left(k,t\right)|^2\simeq4\left|\frac{\lambda\left(\left|k\right|\right)}{\hbar\left(c\left|k\right|-\omega_0\right)}\right|^2\left|f\left(t=0\right)\right|^2\sin^2\left[\left(c\left|k\right|-\omega_0\right)\frac{t}{2}\right]
\end{equation*}
which can be considered to go to 0 when $t$ is large enough, that is, when $(c\left|k\right|-\omega_0)>2\pi/t$. This means that only the modes that vibrate at  ``nearly'' the same frequency as that of the internal mode get coupled to it by the Hamiltonian (\ref{eq:hamham}). One can then consistently assume that $\lambda\left(\left|k\right|\right)$ is constant and takes the values $\lambda\left(\omega_0/c\right)$ because the evolution does not couple the internal mode to external modes which differ too much in energy. Coming back to this property at the end of the treatment it is easy to check that it is expressed by the weak coupling condition $\omega_0\gg\Gamma$, where $\Gamma$ will be introduced as the decay constant. This allows us to write, starting from (\ref{betab}) (see, \emph{e.g.}, \cite{CourbageKaon} for a more rigorous treatment)
\begin{align}
\frac{\mathrm{d}f(t)}{\mathrm{d}t}&=-\int\mathrm{d}k\,\left|\frac{\lambda\left(\left|k\right|\right)}{\hbar}\right|^2\int_0^t\mathrm{d}t'\,\mathrm{e}^{\mathrm{i}\left(\omega_0-c\left|k\right|\right)\left(t-t'\right)}f\left(t'\right)\nonumber\\
&\simeq-\left|\frac{\lambda\left(\frac{\omega_0}{c}\right)}{\hbar}\right|^2\int_0^t\mathrm{d}t'\int\mathrm{d}k\,\mathrm{e}^{\mathrm{i}\left(\omega_0-c\left|k\right|\right)\left(t-t'\right)}f\left(t'\right)\nonumber\\
&=-\left|\frac{\lambda\left(\frac{\omega_0}{c}\right)}{\hbar}\right|^2\int_0^t\mathrm{d}t'\,2\pi\,\delta\left(t-t'\right)f\left(t'\right)\nonumber\\
&=-\frac{\pi}{c}\left|\frac{\lambda\left(\frac{\omega_0}{c}\right)}{\hbar}\right|^2f\left(t\right) \label{eq:Compare}
\end{align}
where the second step is approximately valid on energy conservation grounds. Under these assumptions, $f$ appears as the solution of the first-order differential equation (\ref{eq:Compare}). Together with (\ref{eq:Alpha}) this motivates the following Wigner-Weisskopf exponential ansatz (with $\Lambda$ a complex-valued number):
\begin{equation}
\alpha\left(t\right)=\mathrm{e}^{-\mathrm{i}\omega_0t}\mathrm{e}^{-\frac{1}{2}\Lambda t}.
\end{equation}
This leads to (following a computation very similar to the one found in ref.~\cite{Englert}, sect.~2.5):
\begin{equation*}
\frac{1}{2}\Lambda=\mathrm{i}\int\mathrm{d}k\,\frac{\left|\lambda\left(\left|k\right|\right)\right|^2}{\hbar^2}\,\frac{1-\mathrm{e}^{-\mathrm{i}\left(c\left|k\right|-\omega_0\right)t}\mathrm{e}^{\frac{1}{2}\Lambda t}}{\omega_0-c\left|k\right|-\frac{\mathrm{i}}{2}\Lambda}.
\end{equation*}

We now split $\Lambda$ into its real and imaginary parts: $\Lambda\equiv\Gamma+2\mathrm{i}\omega_{\mathrm{LS}}$, where $\Gamma$ appears to be the transition rate of the decay process, while $\omega_{\mathrm{LS}}$ is the so-called Lamb shift of the (real part of the) energy.

It is shown in \cite{Englert} that in the $\Lambda\rightarrow0^+$ limit (which corresponds to the assumption that the decay takes place over times scales much larger than $\omega_0^{-1}$), we get
\begin{equation} \label{eq:DecayCst}
\Gamma\equiv\frac{2\pi}{c}\frac{\left|\lambda\left(\frac{\omega_0}{c}\right)\right|^2}{\hbar^2}
\end{equation} 
and
\begin{equation} \label{eq:Silence}
\omega_{\mathrm{LS}}\equiv\frac{1}{c\hbar^2}\,\mathrm{vp}\int\mathrm{d}k\frac{\left|\lambda\left(\left|k\right|\right)\right|^2}{c\left|k\right|-\omega_0},
\end{equation}
where $\mathrm{vp}$ denotes the Cauchy principal value of the subsequent integral. Finally, we obtain the following expressions for $\alpha$ and $\beta\left(k,\cdot\right)$:
\begin{subequations} \label{eq:WWWin}
\begin{align}
\alpha\left(t\right)&=\mathrm{e}^{-\mathrm{i}\left(\omega_0+\omega_{\mathrm{LS}}\right)}\mathrm{e}^{-\frac{1}{2}\Gamma t}, \label{eq:WinAlpha}\\
\hspace{-20pt}\beta\left(k,t\right)&=\mathrm{e}^{-\mathrm{i}c\left|k\right|t}\frac{\lambda\left(\left|k\right|\right)}{\hbar}\frac{1-\mathrm{e}^{-\mathrm{i}\left(\omega_0+\omega_{\mathrm{LS}}-c\left|k\right|\right)t}\mathrm{e}^{-\frac{1}{2}\Gamma t}}{c\left|k\right|-\left(\omega_0+\omega_{\mathrm{LS}}\right)+\frac{\mathrm{i}}{2}\Gamma}. \label{eq:WinBeta}
\end{align}
\end{subequations}
The Lamb shift is usually a very small quantity and can be consistently neglected, in which case we recover the same result as in (\ref{eq:Compare}). 

\subsection{Wavefront propagation} \label{sec:Propag}

At this level, it is instructive to consider the situation in which an elementary excitation (single photon) is initially present inside the cavity, with no photon outside. The quantum jump operator commutes with the total photon number operator 
\begin{equation} \label{eq:Number}
\hat{N}\equiv\hbar\omega_0\,\hat{a}_{\mathrm{S}}^\dagger\,\hat{a}_{\mathrm{S}}\otimes\hat{\idmx}_{\mathrm{out}}+\hat{\idmx}_{\mathrm{in}}\otimes\int\mathrm{d}k\,\hat{b}_{\mathrm{S}}^\dagger\left(k\right)\hat{b}_{\mathrm{S}}\left(k\right)
\end{equation}
so that at all times the state remains a single-photon state. In particular, if at time $t=0$, the initial state is located inside the cavity:
\begin{equation} \label{eq:OneInStart}
\mid\!\Psi\left(t=0\right)\rangle=\mid\!1_{\mathrm{in}}\rangle\otimes\mid\!0_{\mathrm{out}}\rangle=\hat{a}^{\dagger}_{\mathrm{S}}\mid\!0_{\mathrm{in}}\rangle\otimes\mid\!0_{\mathrm{out}}\rangle
\end{equation}
then at time $t$ is will still be a coherent superposition of single-photon states in the in and out regions:
\begin{align*}
\mid\!\Psi\left(t\right)\rangle&=\hat{a}_{\mathrm{H}}^\dagger\left(t\right)\mid\!0_{\mathrm{in}}\rangle\otimes\mid\!0_{\mathrm{out}}\rangle\\
&=\alpha^*\left(t\right)\hat{a}^{\dagger}_{\mathrm{S}}\mid\!0_{\mathrm{in}}\rangle\otimes\mid\!0_{\mathrm{out}}\rangle\\
&\hspace{75pt}+\mid\!0_{\mathrm{in}}\rangle\otimes\!\!\int\mathrm{d}k\,\beta^*(k,t)\hat{b}^{\dagger}_{\mathrm{S}}(k)\mid\!0_{\mathrm{out}}\rangle\\
&=\alpha^*\left(t\right)\mid\!1_{\mathrm{in}}\rangle\otimes\mid\!0_{\mathrm{out}}\rangle+\beta^*\left(t\right)\mid\!0_{\mathrm{in}}\rangle\otimes\mid\!1_{\mathrm{out}}\rangle
\end{align*}
where we wrote, symbolically,
\begin{align} \label{eq:Symbol}
\beta^*\left(t\right)\hat{b}_{\mathrm{S}}&\equiv\int\mathrm{d}k\,\beta^*\left(k,t\right)\hat{b}^\dagger_{\mathrm{S}}\left(k\right).
\end{align}
Such states admit a natural interpretation in terms of the photodetection probability function established by Glauber, who showed that the probability per unit of time that a detector located in position $x$, in the ``out'' region, at time $t$ will fire is equal to the modulus square of the single photon amplitude defined as
\begin{equation} \label{eq:Glauber}
\psi\left(x,t\right)\equiv\langle0_{\mathrm{out}}\!\mid\int\mathrm{d}q\,\hat{b}_{\mathrm{S}}\left(q\right)\mathrm{e}^{-\mathrm{i}qx}\int\mathrm{d}k\,\beta^*\left(k,t\right)\hat{b}^{\dagger}_{\mathrm{S}}\left(k\right)\mid\!0_{\mathrm{out}}\rangle
\end{equation}
as is shown in \cite{FoxMulder} (sect.~4.2) in the case of spontaneous emission of a two-level atom in free, three-dimensional space. 

We now use the results of sect.~\ref{sec:WigWei} to derive an expression for the spacetime dependence of the Glauber wavepacket in the outside region. Since we consider single-photon state s we might call this object the ``photonic wave function''. Here we want to see how the wavepacket corresponding to an initially confined single-photon state evolves. Hence we look at the quantities (we assume for simplicity that all modes inside and outside are characterized by a same polarisation):
\begin{align*}
\psi_{\mathrm{rg/lf}}\left(x,t\right)&=\idmx_{\mathrm{rg/lf}}\left(x\right)\langle0_{\mathrm{out}}\!\mid\int_a^b\mathrm{d}k\,\beta^*\left(k,t\right)\hat{b}_{\mathrm{S}}\left(k\right)\,\mathrm{e}^{-\mathrm{i}kx}\mid\!k_{\mathrm{out}}\rangle\\
&=\idmx_{\mathrm{rg/lf}}\left(x\right)\int_a^b\mathrm{d}k\,\beta^*\left(k,t\right)\,\mathrm{e}^{-\mathrm{i}kx}
\end{align*}
where $\idmx_{\mathrm{rg}}$ and $\idmx_{\mathrm{lf}}$ are indicator functions for the right and left regions respectively (see sect.~\ref{sec:QJO} and equation (\ref{eq:Oliver}). In these integrals we have $a=0$ and $b=+\infty$ for right-outgoing waves, and $a=-\infty$ and $b=0$ for left-outgoing waves.

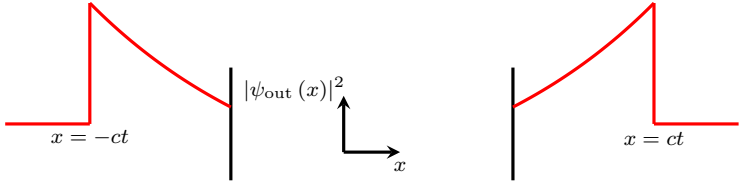
\begin{figure}[t]
\begin{center}
\begin{tikzpicture}[very thick, >=stealth, scale=.75]
\draw[->] (-.5,-.5) -- (.5,-.5);
\draw (.5,-.75) node {\scriptsize $x$};
\draw[->] (-.5,-.5) -- (-.5,.5);
\draw (-1.375,.625) node {\scriptsize $\left|\psi_{\mathrm{out}}\left(x\right)\right|^2$};
\draw (-2.5,-1) -- (-2.5,1);
\draw (2.5,-1) -- (2.5,1);
\draw[red] (-6.5,0) -- (-5,0);
\draw[red] (-5,0) -- (-5,2.145);
\draw[domain=-5:-2.5][red] plot(\x,{exp(-\x/3.75)-exp(.5)});
\draw[domain=2.5:5][red] plot(\x,{exp(\x/3.75)-exp(.5)});
\draw[red] (5,2.145) -- (5,0);
\draw[red] (5,0) -- (6.5,0);
\draw (-5,-.25) node {\scriptsize $x=-ct$};
\draw (5,-.25) node {\scriptsize $x=ct$};
\end{tikzpicture}
\end{center}
\vspace{-10pt}
\caption{Profile of the wave function square modulus in the outside region at fixed time. It increases exponentially with increasing $\pm x$ (see text and (\ref{eq:Lightcone})) and vanishes outside the light cone.\label{Wavefront}}
\end{figure}

From the expression (\ref{eq:WinBeta}) of $\beta\left(k,\cdot\right)$ we get (we make the usual \cite{FoxMulder,Fermi} approximation of extending the integration domain to the whole real line because the main contribution is centered around the pole which lies close to the positive real axis and far from the origin), introducing the notation $\bar{\omega_0}\equiv\omega_0+\omega_{LS}$:
\begin{subequations} \label{eq:IntheDark}
\begin{align}
\hspace{-20pt}\psi_{\mathrm{rg}}\left(x,t\right)&=\idmx_{\mathrm{rg}}\left(x\right)\int\mathrm{d}k\,\frac{\lambda^*\left(\left|k\right|\right)}{ck-\bar{\omega}_0-\frac{\mathrm{i}}{2}\Gamma}\nonumber\\
&\hspace{25pt}\left[\mathrm{e}^{-\mathrm{i}k\left(x-ct\right)}-\mathrm{e}^{\mathrm{i}\left(\bar{\omega}_0t-kx\right)}\mathrm{e}^{-\frac{1}{2}\Gamma t}\right]\label{eq:right},\\
\hspace{-20pt}\psi_{\mathrm{lf}}\left(x,t\right)&=\idmx_{\mathrm{lf}}\left(x\right)\int\mathrm{d}k\,\frac{\lambda^*\left(\left|k\right|\right)}{-ck-\bar{\omega}_0-\frac{\mathrm{i}}{2}\Gamma}\nonumber\\
&\hspace{25pt}\left[\mathrm{e}^{-\mathrm{i}k\left(x+ct\right)}-\mathrm{e}^{\mathrm{i}\left(\bar{\omega}_0t-kx\right)}\mathrm{e}^{-\frac{1}{2}\Gamma t}\right]\label{eq:left}.
\end{align}
\end{subequations}
To compute these integrals, we use the Cauchy residue theorem. The calculation (for details see \cite{FoxMulder}) yields
\begin{equation} \label{eq:Lightcone}
\hspace{-5pt}\psi_{\mathrm{rg/lf}}\left(x,t\right)=\pm\frac{\mathrm{2i\pi}}{\hbar}\idmx_{\mathrm{rg/lf}}\left(x\right)\!\Theta\!\left(ct\mp x\right)\!\lambda^*\!\left(\frac{\bar{\omega}_0}{c}\right)\mathrm{e}^{\frac{\mathrm{i}}{c}\left(\bar{\omega}_0+\frac{\mathrm{i}}{2}\Gamma\right)\left(ct\mp x\right)}
\end{equation}
where $\Theta$ is the usual Heaviside step function. Thus we see that the wave function is nonvanishing inside the light cone only. At fixed time, it increases exponentially -within the light cone- with increasing $x$ or $-x$, depending on whether we look at what happens on the right or on the left of the cavity (see Fig.~\ref{Wavefront}). As we see photons are located at a distance of the order of $c/\Gamma$ where $\Gamma$ is the inverse of the lifetime of the mode inside the cavity. The spatial extent of the photonic wave packet is also of the order of $c/\Gamma$, in agreement with time-energy uncertainty. Although it is, as far as we know, not common, in the Friedrichs formalism, to give an explicit spacetime dependence for the decay products, this has been done in quantum optics \cite{FoxMulder,Brainis} and also in the leaky mode treatment of Maxwell's equations \cite{BrillouinWigner}. 

\section{On the time evolution of coherent states} \label{sec:Coherent}

In cavity QED experiments, photons are kept in what is coloquially referred to as a box, and which is in fact a superconducting cavity during so-called macroscopic time intervals (130 milliseconds \cite{HarocheNature} in the case of Serge Haroche's group), after which they are absorbed by the cavity walls. We shall now use the model defined in sect.~\ref{sec:Jump} to study the behavior of coherent states in QED cavities, in particular their status of ``classical pointer states''.

\subsection{Decoherence and pointer states} \label{sec:Decoh}

The decoherence program is aimed, among others, at explaining why the macroscopic world does not (directly) obey quantum mechanical laws but instead behaves classically \cite{ZurekToday}. It is based on the insight that quantum systems are never isolated \cite{Schlosshauer}, but that their evolution is at least partly monitored by their environment with which they continuously interact (as is put in \cite{HarocheRaimond} (sect.~4), ``the environment is watching''). The goal of the decoherence program is to build on this insight to explain how the ``classical'' properties of the macroscopic world can be inferred from a purely quantum mechanical paradigm at the micropscopic level. In this approach, one can show \cite{ZurekPRD,ZurekToday} how the interaction between the system and its environment destroys the coherence between a paticular set of quantum states, the so-called ``pointer states'', which correspond to the ``pointer'' positions of the macroscopic measurement apparatus, \emph{i.e.}, the different outcomes of a macroscopic measurement performed on the system \cite{Schlosshauer}. The so-called quantum pointer states are ``quasi-classical'' states of a quantum system. They are singled out by their robustness to the interaction between the system and its environment. Since decoherence is, as is well-known, the corollary of entanglement \cite{Article,Cosmos}, pointer states of the system can be defined as the states which become minimally entangled with the environment in the course of their evolution \cite{ZurekNature}. In our model, the considered system is the cavity itself while the outside region plays the role of the environment.

\subsection{Coherent states as pointer states of the cavity} \label{sec:Bose}

In this section we show that coherent states of the cavity are pointer states. It is indeed possible, as we do, to show that they do not get entangled with the environment (that is, the outside region) in the course of their evolution. To do that, we first look at how a $n$-photon cavity Fock state evolves. Staying in the Heisenberg picture, we are thus interested in the evolution of
\begin{align*}
&\mid\!\varphi_n\left(t=0\right)\rangle=\frac{1}{\sqrt{n!}}\left(\hat{a}_{\mathrm{S}}^\dagger\right)^n\mid\!0_{\mathrm{in}}\rangle\otimes\mid\!0_{\mathrm{out}}\rangle\\
\stackrel{t}{\rightarrow}&\mid\!\varphi_n\left(t\right)\rangle=\frac{1}{\sqrt{n!}}\left(\hat{a}_{\mathrm{H}}^\dagger\left(t\right)\right)^n\mid\!0_{\mathrm{in}}\rangle\otimes\mid\!0_{\mathrm{out}}\rangle.
\end{align*}
Let us write symbolically
\begin{subequations}
\begin{align}
\beta\left(t\right)\hat{b}_{\mathrm{S}}&\equiv\int\mathrm{d}k\,\beta\left(k,t\right)\hat{b}_{\mathrm{S}}\left(k\right),\\
\mid\!n_{\mathrm{out}}\rangle&\equiv\frac{1}{\sqrt{n!}}\left(\hat{b}_{\mathrm{S}}^\dagger\right)^n\mid\!0_{\mathrm{out}}\rangle
\end{align}
\end{subequations}
so that we can expand $\mid\!\varphi_n\left(t\right)\rangle$ more conveniently. Using (\ref{eq:Ansatz}) we get (since $\hat{a}_{\mathrm{S}}^\dagger$ and $\hat{b}_{\mathrm{S}}^\dagger$ commute)
{\allowdisplaybreaks
\begin{align*}
\mid\!\varphi_n\left(t\right)\rangle&=\frac{1}{\sqrt{n!}}\left(\alpha^*\left(t\right)\hat{a}_{\mathrm{S}}^\dagger+\beta^*\left(t\right)\hat{b}_{\mathrm{S}}^\dagger\right)^n\mid\!0_{\mathrm{in}}\rangle\otimes\mid\!0_{\mathrm{out}}\rangle\\
&=\frac{1}{\sqrt{n!}}\sum_{k=0}^n\frac{n!}{k!\left(n-k\right)!}\left(\alpha^*\left(t\right)\right)^k\left(\beta^*\left(t\right)\right)^{n-k}\\
&\hspace{75pt}\left(\hat{a}_{\mathrm{S}}^\dagger\right)^k\left(\hat{b}_{\mathrm{S}}^\dagger\left(t\right)\right)^{n-k}\mid\!0_{\mathrm{in}}\rangle\otimes\mid\!0_{\mathrm{out}}\rangle\\
&=\sum_{k=0}^n\sqrt{\frac{n!}{k!\left(n-k\right)!}}\left(\alpha^*\left(t\right)\right)^k\left(\beta^*\left(t\right)\right)^{n-k}\\
&\hspace{75pt}\mid\!k_{\mathrm{in}}\rangle\otimes\mid\!\left(n-k\right)_{\mathrm{out}}\rangle.
\end{align*}
}
Thus if at time $t=0$ the cavity is prepared in a coherent state and the outside is in its vacuum state, then the evolution generates a tensor product of a cavity coherent state with a coherent state of the outside:
\begin{align} \label{eq:Evocoh}
\mid\!\psi\left(t=0\right)\rangle&=\mathrm{e}^{-\frac{\left|\xi\right|^2}{2}}\sum_{n=0}^{+\infty}\frac{\xi^n}{\sqrt{n!}}\mid\!n_{\mathrm{in}}\rangle\otimes\mid\!0_{\mathrm{out}}\rangle\nonumber\\
\rightarrow\mid\!\psi\left(t\right)\rangle&=\mathrm{e}^{-\frac{\left|\xi\right|^2}{2}\left|\alpha\left(t\right)\right|^2}\sum_{n=0}^{+\infty}\sum^{n}_{m=0}\frac{\left(\xi\alpha^*\left(t\right)\right)^m}{\sqrt{m!}}\mid\!m_{\mathrm{in}}\rangle\nonumber\\
&\otimes\mathrm{e}^{-\frac{\left|\xi\right|}{2}\left|\beta\left(t\right)\right|^2}\frac{\left(\xi\beta^*\left(t\right)\right)^{n-m}}{\sqrt{\left(n-m\right)!}}\mid\!\left(n-m\right)_{\mathrm{out}}\rangle
\end{align}
where we used the identity $\left|\alpha^*\left(t\right)\right|^2+\left|\beta^*\left(t\right)\right|^2=1$. The state at time $t$ can be rewritten as 
\begin{align} \label{eq:Coherent}
\mid\!\psi\left(t\right)\rangle&=\mathrm{e}^{-\frac{\left|\xi\right|^2}{2}\left|\alpha\left(t\right)\right|^2}\sum_{k=0}^{+\infty}\frac{(\xi\alpha^*\left(t\right))^{k}}{\sqrt{k!}}\mid\!k_{\mathrm{in}}\rangle\nonumber\\
&\otimes\mathrm{e}^{-\frac{\left|\xi\right|^2}{2}\left|\beta\left(t\right)\right|^2}\sum_{m=0}^{+\infty}\frac{(\xi\beta^*\left(t\right))^m}{\sqrt{m!}}\mid\!m_{\mathrm{out}}\rangle.
 \end{align}
This establishes that coherent states of the cavity interact with the exterior without getting entangled with it. They can thus be considered as ``classical pointers'' according to the criterion for classicality derived by Zurek in \cite{ZurekNature}. This result is similar to the one that exists for beam splitters (see, for instance, \cite{Loudon} (sect.~5.9)). A common argument for the classicality of coherent states is based on the Lindblad master equation, which is a widely used dissipative (\emph{i.e.} nonunitary) equation which models the coherence loss which a quantum system undergoes when it interacts with its environment. It governs the studied system's reduced density matrix, and, in the case of a QED cavity (at zero temperature), it reads \cite{HarocheRaimond} (sect.~7.5.1)
\begin{align} \label{eq:Master}
\frac{\mathrm{d}}{\mathrm{d}t}\hat{\rho}_{\mathrm{in}}\left(t\right)&=\frac{1}{\mathrm{i}\hbar}\left[\hat{H}_{\mathrm{in}},\hat{\rho}_{\mathrm{in}}\left(t\right)\right]\nonumber\\
&+\frac{\Gamma}{2}\left[2\hat{a}\hat{\rho}_{\mathrm{in}}\left(t\right)\hat{a}^\dagger-\hat{a}^\dagger\hat{a}\hat{\rho}_{\mathrm{in}}\left(t\right)-\hat{\rho}_{\mathrm{in}}\left(t\right)\hat{a}^\dagger\hat{a}\right].
\end{align}
In \cite{Article} we proposed an alternate derivation of the Lindblad master equation. Our proof is based on the fact that any damped coherent state of the form 
\begin{equation} \label{eq:Anycoh}
\mid\!\xi_{\mathrm{in}}\left(t\right)\rangle=\mathrm{e}^{-\frac{\left|\xi\right|^2}{2}\left|\alpha\left(t\right)\right|^2}\sum_{m=0}^{+\infty}\frac{\left(\xi\alpha\left(t\right)\right)^m}{\sqrt{m!}}\mid\!m_{\mathrm{in}}\rangle,
\end{equation}
with $\alpha$ given by (\ref{eq:WWWin}) and $\hat{H}=\hbar\omega_0\,\hat{a}^\dagger\hat{a}$, obeys the Lindblad equation (\ref{eq:Master}). Since coherent states constitute a basis of the Hilbert space (rigorously, they form an overcomplete basis), we find that, by virtue of the linearity of the master equation, any state is a solution of the Lindblad master equation. The proof given in \cite{Article} made resort to first quantization arguments. The derivation of (\ref{eq:Coherent}) given here is however more traditional because it does only appeal to standard second quantization techniques.

\section{Classical analysis of a QED cavity} \label{sec:Maxwell}

\subsection{Damped coherent states and the classical limit} \label{sec:Ehrenfest}

It is sometimes argued that the Ehrenfest theorem establishes that quantum average values behave like classical quantities. This is wrong because, even if one can, in a variety of situations, derive an equation that looks classical at first sight, that is, an equation of the type $m\,\mathrm{d}^2\left<\mathbf{r}\right>/\mathrm{d}t^2=\left<\mathbf{F}\right>$, this equation is not equivalent to its classical counterpart in general. For instance in the case where the force derives from a potential, $\mathbf{F}=-\nabla V\left(\mathbf{r},t\right)$ and $\left<\mathbf{F}\right>=-\left<\nabla V\left(\mathbf{r},t\right)\right>$, but, generically, $\left<\nabla V\left(\mathbf{r},t\right)\right>\neq\nabla V\left(\left<\mathbf{r}\right>,t\right)$ and it is only in situations where we can neglect quantum fluctuations that classical mechanics can be inferred from quantum mechanics, which is in a sense a tautology. The damped harmonic oscillator however constitutes an exception in the sense that regardless of the actual state the oscillator finds itself in, we have $m\,\mathrm{d}^2\langle x\rangle/\mathrm{d}t^2=F\left(\langle x\rangle\left(t\right),\mathrm{d}\langle x\rangle/\mathrm{d}t\right)$. This is easily established by computing the time evolution of $\left<\hat{a}\right>$ using the master equation (\ref{eq:Master}). One finds that
\begin{equation} \label{eq:awithtime}
\frac{\mathrm{d}\left<\hat{a}\right>}{dt}=\left(-\mathrm{i}\omega-\frac{\Gamma}{2}\right)\left<\hat{a}\right>\left(t\right).
\end{equation}
After integration over time we get
\begin{equation*}
\left<\hat{a}\right>\left(t\right)=\left<\hat{a}\right>\left(t=0\right)\mathrm{e}^{\left(-\mathrm{i}\omega-\frac{\Gamma}{2}\right)t}.
\end{equation*}
Making use of the fact that $\left<x\right>\left(t\right)=\left<\hat{a}\right>\left(t\right)+\left<\hat{a}\right>^*\left(t\right)$, we get finally that $\left<x\right>(t)=A\cos\left(\omega t+\phi_0\right)\exp\left(\left(-\Gamma/2\right)t\right)$, which is the most general solution of the dynamical equation of a classical-one dimensional-damped oscillator (that is, $\mathrm{d}^2x/\mathrm{d}t^2+\Gamma\,\mathrm{d}x/\mathrm{d}t+\tilde{\omega}^2x=0$, with $\tilde{\omega}^2\equiv\omega^2+\Gamma^2/4$). From this point of view, it is justified to describe the average behavior of a damped oscillator trapped in a QED cavity  as a classical oscillator, and thus to resort to the Maxwell equations in order to study its properties, as we shall do in the next section. It is by the way particularly relevant to approximate the behavior of the quantum oscillator by its average behavior in the case of coherent states because they minimize Heisenberg uncertainties, which emphasizes their status of ``quasi-classical states''. They also remain disentangled with the environment which renders them classical pointer states in Zurek's sense \cite{ZurekNature,Article}.

\subsection{Quasimodal analysis of a QED cavity} \label{sec:Maxwell}

In this section we study the open QED cavity described in \cite{Pipeau} by searching for its eigenmodes and complex eigenfrequencies, using a Finite Element Method (FEM). This classical electrodynamic approach allow us to derive a number of features observed experimentally by Haroche and coworkers (\cite{HarocheNature}).

\subsubsection{Searching resonances of open resonators using perfectly matched layers} \label{sec:Eigen}

The major difficulty in the treatment of open problems in a numerical scheme based on a finite computational window is to deal with infinity issues. Since their introduction by Bérenger in \cite{Berenger1994185} for the time dependent Maxwell's equations, Perfectly Matched Layers (PMLs) have become a widely
 used technique in computational physics. The idea is to enclose the area of interest by surrounded layers
which are absorbing and perfectly reflectionless. These absorbing boundary conditions can be understood in the global framework of transformation optics (\cite{Nicolet}). The principle of the technique is to perform a geometrical transformation (here a complex stretch of coordinates), leading to equivalent material properties (\cite{FPCF,Lassas2001739,CambridgeJournals:1202020}).\\

The spectral problem we are dealing with consists in finding the solutions of source free Maxwell's equations, \textit{i.e.} finding complex eigenvalues $\Lambda_n=(\omega_n/c)^2$ and non zero eigenvectors $\mathbf{E}_n$  such that :

\begin{equation}
\mathcal{M}(\mathbf{E}_n):=\nabla\times\left(\tens{\mu}^{-1}\cdot\nabla\times\mathbf{E}_n\right)=\Lambda_n \,\tens{\varepsilon}\cdot\mathbf{E}_n.
\label{eq:eigenpb}
\end{equation}
where $\tens{\varepsilon}$ and $\tens{\mu}$ are the relative dielectric permittivity and magnetic permeability tensors describing the electromagnetic properties of the system (cavity+external world).\\
For Hermitian open problems, the generalized spectrum of Maxwell's operator $\mathcal{M}$ is real and composed of two parts : the discrete spectrum with trapped modes exponentially decreasing at infinity, and the continuous spectrum with radiation modes oscillating at infinity. In addition, another type of solution is present and very useful to characterize the spectral properties of unbounded structures : the
so-called \textit{leaky modes} (also termed quasimodes, quasi normal modes (\cite{FPCF,Settimi2009}) or
quasi guided modes (\cite{Tikhodeev2002}) in the literature). These eigenmodes with complex associated frequency are an intrinsic feature of open waveguides.\\
PMLs have proven to be a very convenient tool to compute leaky modes in different configurations (\cite{Zolla:07,HEIN2004}). Indeed they efficiently mimic the infinite space provided a suitable choice of their parameters.
The introduction of infinite PMLs rotates the continuous spectrum in the complex plane (since the operator involved in the problem is now a non self-adjoint extension of the original self-adjoint operator). The effect is not only to turn the continuous spectrum energies into complex valued-quantities but it also unveils the leaky modes in the region swept by the rotation of this essential spectrum (\cite{thesebenjg}). It is important to note that leaky modes do not depend on the choice of a particular complex stretching : adding the PMLs is only a way to discover them. Finally, in order to apply the FEM, the PMLs have to be truncated at finite distance which results in an operator having only point spectrum with approximate radiation modes (also termed as PML modes or Bérenger modes) due to the discretization of the continuous spectrum by finite PMLs (\cite{olyslager:1408}). Actually, the PML technique is analog to the theory of analytic dilation (\cite{Aguilar1971, Balslev1971}) developed in the 1970s for Schrödinger Hamiltonians.\\
In the following, eigenvalues are denoted $\omega_n=\omega_n'+\mathrm{i}\omega_n''$. The real part is the resonant angular frequency $\omega'_n=2\pi f_n$ and the imaginary part is the damping coefficient, which is related to the lifetime $\tau_n$ of the photon in the cavity by $\omega_n''=2\pi/\tau_n$. The quality factor associated to a resonance is defined by  $Q_n=\omega_n'/(2\omega_n'')$.

\subsubsection{Numerical set up} \label{sec:Simu}

A detailed description of the cavity used by Haroche and woworkers can be found in \cite{Pipeau}. 
It is composed of two mirrors of diameter $D=50\mm$ facing 
each other. The distance between their apexes is $L=27.57\mm$, 
and their surface is toroidal with radii of curvature $r=39.4\mm$ in 
the $Oxz$ plane and $R=40.6\mm$ in the $Oyz$ plane. The mirrors are 
coated with a thick layer of superconducting niobium. 
We take advantage of the symmetries of the problem and thus model only one eighth of the 
cavity to save memory and computation time. By setting a well chosen combination of Neumann 
and Dirichlet boundary conditions on the cutting planes, we can select the modes with desired 
symmetries. The eigenproblem defined by (\ref{eq:eigenpb}) is then solved by the FEM, using 
cartesian PMLs terminated by homogeneous Neumann boundary conditions to truncate the infinite 
space. The computational cell is meshed using second order edge elements, with 
a maximum size of an element set to $\lambda^{\rm r}/\left(|\re\,\varepsilon|N\right)$, 
where $\lambda^{\rm r}=3.68\cm$ is the approximate resonant wavelength 
of the cavity, and $N$ is an integer ($N=7$ for the domain inside of the cavity, $N=7$ 
for the domain outside the cavity, $N=5$ for the PMLs and $N=N_{\rm m}$ for the mirror 
surfaces). The final algebraic system is solved using a direct solver (PARDISO).\\  

In a first approximation, both mirrors are assumed to be lossless, thus we set Perfect Electric 
Conductor boundary conditions (PEC, \emph{i.e.} Dirichlet boundary conditions).
In a second step, in order to account for losses, absorption is considered through a 
Surface Impedance Boundary Condition (SIBC) on the boundaries of both mirrors.

\begin{figure}
\begin{center}
{\includegraphics[width=1\columnwidth]{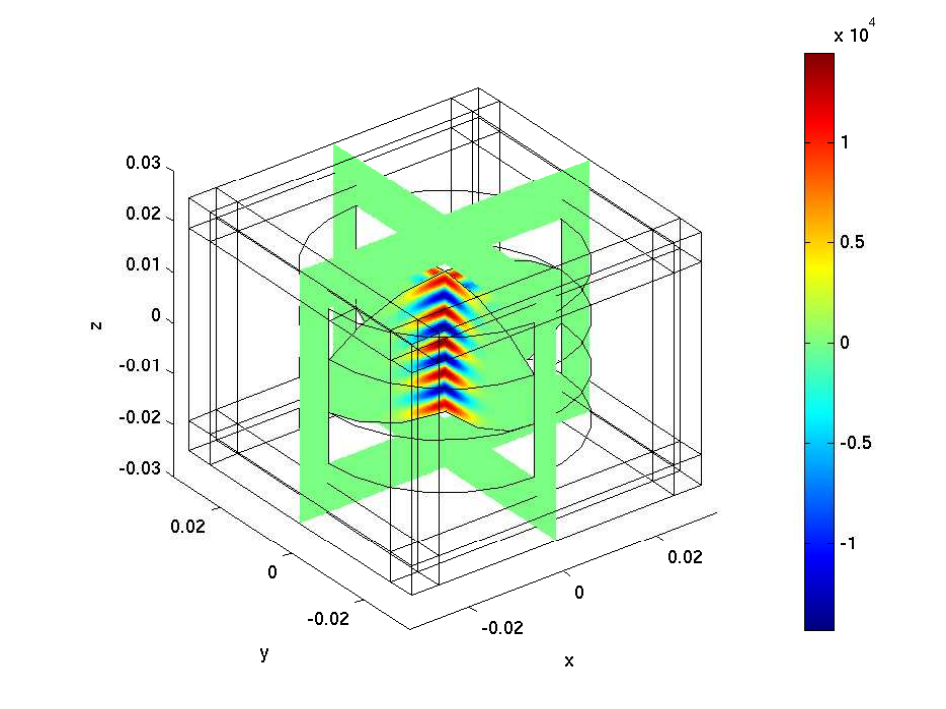}}\subcaption{$\re(E_x)$, mode 1.\label{fig:mode1}}
{\includegraphics[width=1\columnwidth]{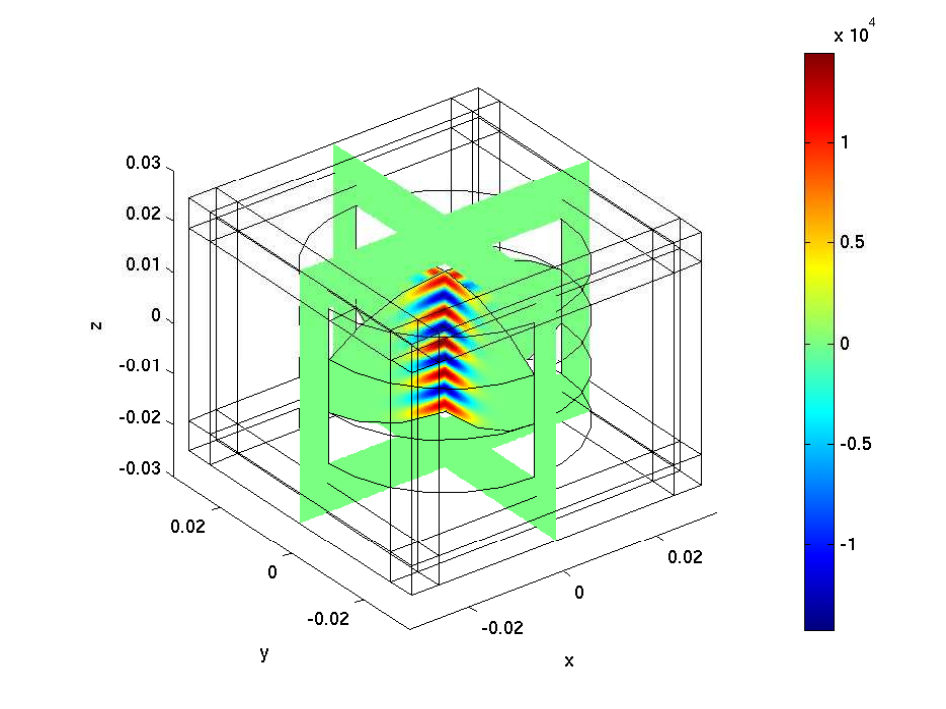}}\subcaption{$\re(E_y)$, mode 2.\label{fig:mode2}}
\end{center}
\caption{Field maps of the two eigenmodes. \label{fig:modes}}
\end{figure}

\subsubsection{A first approach: Perfect mirrors} \label{sec:Perfect}

Let us consider first a lossless model, \emph{i.e.} perfect mirrors. 
Among many others, we find only two eigenfrequencies with exceptionally low imaginary parts 
(or high associated lifetimes). Their real parts correspond to frequencies $f_1$ and $f_2$ located 
around $51.085\GHz$ (to be compared to $51.099\GHz$ found experimentally in 
\cite{Pipeau}), with $f_{12}=f_1-f_2\simeq1.2\MHz$ (same as in \cite{Pipeau}). 
According to the well known Fabry-Pérot interferometry principle, a 
modification of the distance between the mirrors would allow to retrieve the experimental frequency,
as shown in the following section.
The small discrepancy between the two complex eigenfrequencies is due to the 
removed degeneracy induced by the slight cylindrical symmetry breaking of cavity ($r\neq R$). 
The mode labelled 1 with higher frequency is $x$-polarized while mode 2 with 
lower frequency is $y$-polarized, and they both 
have nine antinodes along $Oz$ (modes TEM$_{900}$, see the field maps on Fig.~\ref{fig:modes}). The modes found to have the highest lifetimes are those studied in \cite{Maioli} on the basis of \cite{LaserBeams}.

\begin{figure}
\begin{center}
{\includegraphics[width=1\columnwidth]{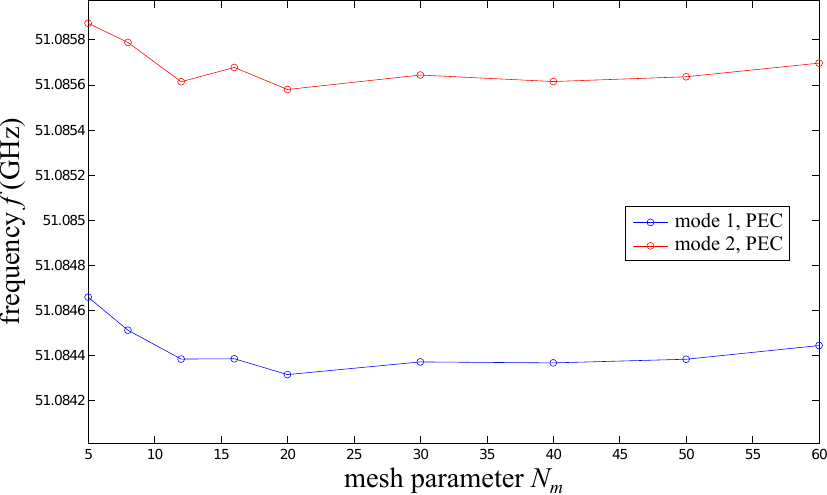}}\subcaption{\label{fig:conv_f}}
{\includegraphics[width=1\columnwidth]{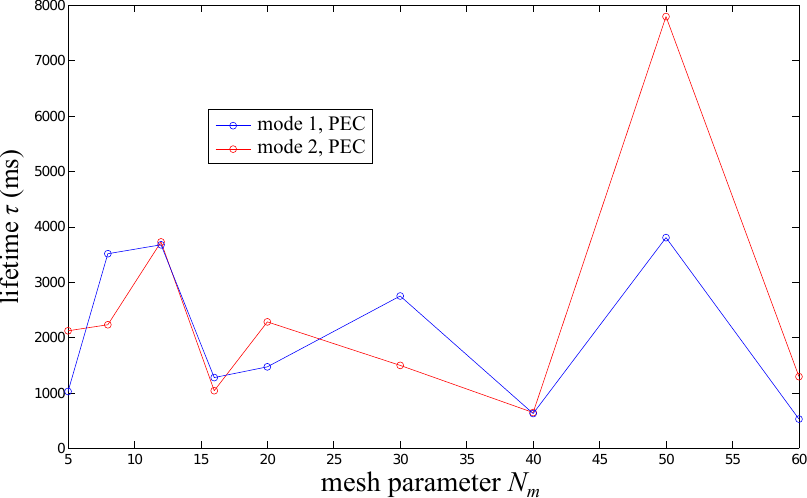}}\subcaption{\label{fig:conv_t}}
\end{center}
\caption{Convergence of the eigenfrequencies with mesh parameter $N_{\rm m}$ for modes 1 and 2: (a) resonant frequency $f$, (b) lifetime $\tau$}
\end{figure}

We study the convergence of the eigenfrequencies as a function of the mesh refinement.
A satisfying convergence is obtained with increasing $N_{\rm m}$
(Fig.~\ref{fig:conv_f}) for the real part of the frequencies, while all that can be said
about lifetimes (Fig.~\ref{fig:conv_t}) is that they have an order of magnitude 
of a few seconds. This non-convergence is due to the fact that the 
resonances features are in that case dominated by the geometry of the mirrors surface, which is 
approximated by tetrahedral elements, thus not exactly reproducing the smooth 
curvature of the toroidal surfaces. To be more specific, Haroche \emph{et al.} 
fabricated two cavities $M_1$ and $M_2$. 
They measured a \emph{fabricated} maximum peak-to-valley deviation 
from the ideal toroidal mirror shape of only $300\nm$. 
We can easily estimate a \emph{numerical} maximum peak-to-valley deviation of 
our virtual tessellated mirrors. The 3D mesh is made of tetrahedra, 
so the surfaces of interest of each mirror are discretized into a set of 
triangles of side lengths set to a maximal value of $\lambda^{\rm r}/\left(|\re\,\varepsilon|N\right)$.
Let us consider an equilateral triangle $T$ of side length 
$\lambda^{\rm r}/\left(|\re\,\varepsilon|N\right)$ 
whose vertices belong to a sphere of radius $r$. When approximating the 
corresponding piece of sphere by this triangle T, the maximum deviation 
from the original shape is the minimum
distance between the barycenter of T and the sphere.
For $N_m=22$, this characteristic length is approximately $300\nm$, 
which corresponds to the experimental value. 
Our numerical mirrors are globally closer from the exact toroidal 
shape than the experimental ones. However, numerically taking into 
account the remaining local surface roughness of typically $10\nm$ 
is not only numerically unthinkable but also irrelevant (details of $\lambda_r$ 
over half a million). Finally, the lossless PEC model represents a 
cavity so resonant that even a tiny change (setting $N_m$ to $60$ leads 
to a maximum deviation from the exact shape of only $40\nm$, \emph{i.e.} $\lambda_r/150000$) in discretization 
leads to a significantly different lifetime. 
Beside the non-convergence of this first model, the order of magnitude of the lifetime is seriously larger than lifetimes found experimentally (seconds instead of $100\ms$ in \cite{Pipeau,HarocheNature}). 
At first sight, assuming a perfect geometry (perfect mirrors with ideal shape and alignment), losses are not dominated by radiation loss.
We shall now try two approaches in order to give an estimate of the effective resistive losses. In the first one (sect.~ \ref{sec:Loss}) , we include some friction mechanism (absorption by the mirrors). As we shall show, from a numerical point of view, 
including some friction mechanism will benefit to the model convergence. However, the absorption intorduced in order to mimic the observed losses is not consistent with measures of the resistivity of the mirrors presented in  \cite{Pipeau}. This apparent paradox brings us to the last possibility which is to explain losses in terms of the rugosity of the surface of the mirrors. We develop in this context a purely geometric model (sect.~ \ref{sec:rugos}), formulated in terms of ray optics which provides a satisfactory agreement with observations.

\subsubsection{An approach based on Lossy Mirrors} \label{sec:Loss}

We now apply Surface Impedance Boundary Conditions (SIBC) 
$Z_{\rm s}=X_{\rm s}+\mathrm{i}Y_{\rm s}$ on the faces of the mirrors.
London penetration depth for niobium $L_{\rm L}$ (independent of frequency) 
is set to a typical value of $0.1\micron$ \cite{rodwell2001high}.
In the framework of a two-fluid model, at low temperature, the imaginary 
(inductive) part of the impedance can be approximated by 
$Y_{\rm s}=\omega\mu_0 L_{\rm L}=6.4\muOhm$. 
As for the real (resistive) part of the impedance, 
it is extremely difficult to measure
and greatly depends on numerous experimental conditions.
Therefore, our only option is to estimate this parameter to 
obtain lifetimes of the same order of magnitude as measured in \cite{Pipeau}.
Finally, we adjusted the length of the cavity to find resonant frequencies closer to 
those measured in \cite{Pipeau}. 

\begin{figure}
\begin{center}
{\includegraphics[width=1\columnwidth]{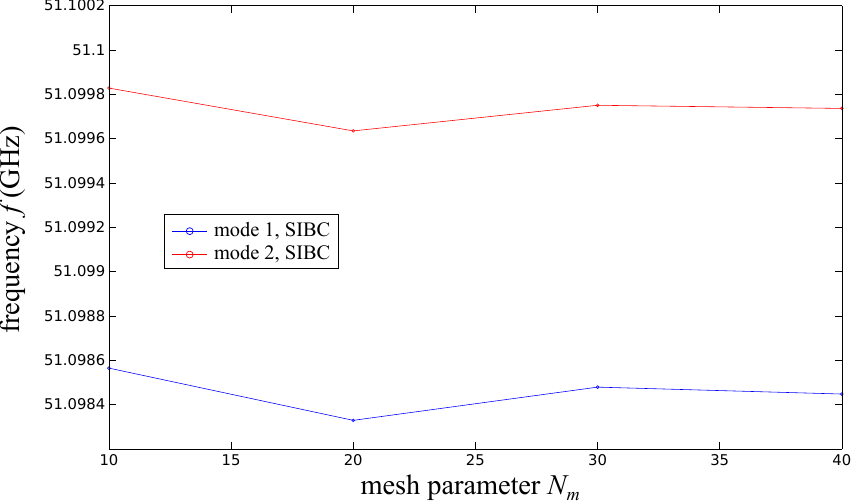}}\subcaption{\label{fig:conv_f2}} 
{\includegraphics[width=1\columnwidth]{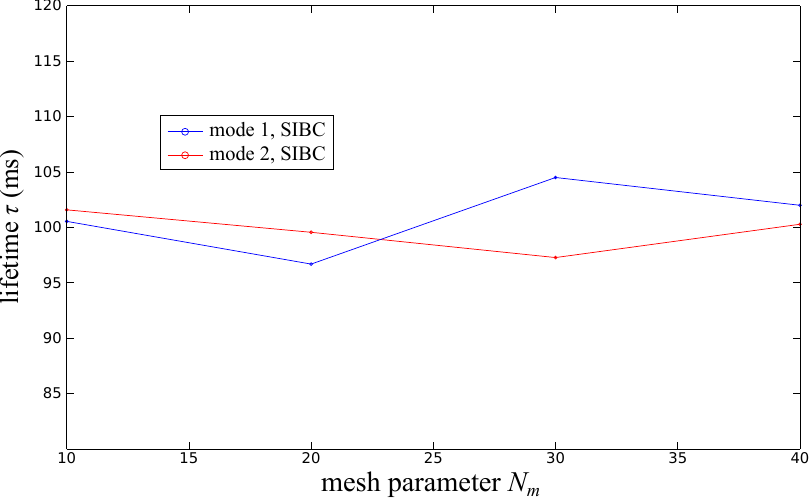}}\subcaption{\label{fig:conv_t2}}
\end{center}
\caption{Convergence of the eigenfrequencies with mesh parameter $N_{\rm m}$ for modes 1 and 2: (a) resonant frequency $f$, (b) lifetime $\tau$}
\end{figure}

With the updated value of $\textrm{L}=27.562\mm$ (instead of
$27.57\mm$ in \cite{Pipeau}), the system exhibits two resonant frequencies
$f_1=51.0984\GHz$ and $f_2=51.0997\GHz$ 
($f_{12}=f_1-f_2\simeq1.29\MHz$). As in the PEC model, the 
convergence as function of mesh refinement is reached (see Fig.~\ref{fig:conv_f2}) for $N_m=40$.

Moreover, the discrepancy between lifetimes when refining the mesh size on the mirror is greatly 
reduced compared to the PEC case (see Fig.~\ref{fig:conv_t2}). 
With $X_{\rm s}=1\muOhm$, we 
obtain lifetimes of $\sim100\ms$ for both modes, which corresponds to 
the average lifetime found in \cite{Pipeau} for both cavities: 
$112\pm4\ms$ (LF mode) and $87\pm10\ms$ (HF mode) for $M_1$,
$74\pm6\ms$ (LF mode) and $130\pm4\ms$ (HF mode) for $M_2$.
This value of $X_{\rm s}$ should be seen as an upper bound of the resistive phenomenon. 
It includes all loss processes beyond radiation loss: roughness of the mirrors, superconductor imperfections, etc. It is worth noting however that resistivity has been measured \cite{Pipeau} at low temperature and is not consistent with the value $X_{\rm s}=1\muOhm$ that we imposed in order to find good agreement with the observed lifetime. Indeed, at working temperature ($0.8\mathrm{ K}$) the resistance of the mirrors is $3\pOhm$, as can be inferred from \cite{Pipeau}. In a sense this is not amazing because closed cavities are characterized by a lifetime of the order of one second \cite{ClosedCavity} which suggests that resistivity of the mirror surface does not suffice to explain the decay of the internal mode of the cavity. There remains one possible source of imperfection that we did not discuss so far, which is the rugosity of the surface. We explore in sect.~\ref{sec:rugos} a simple explanation formulated in terms of ray optics which suffices to qualitatively derive the observed value of the lifetime.

\subsection{A derivation of the lifetime based on rugosity and ray optics.} \label{sec:rugos}

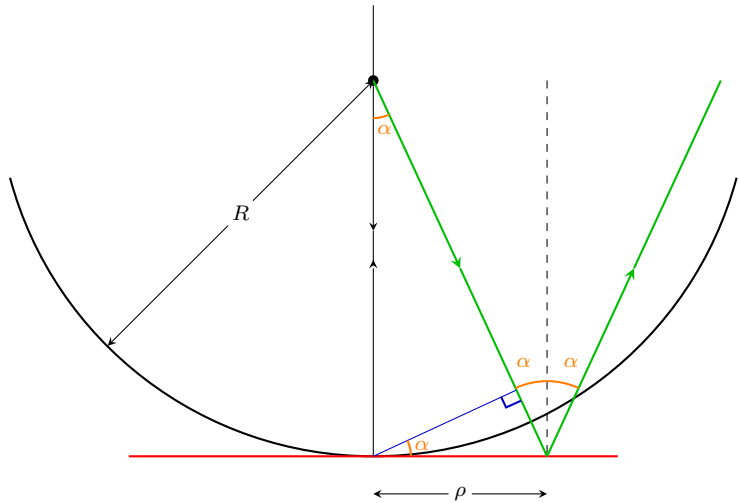
\begin{figure}
\begin{center}
\begin{tikzpicture}[thick, >=stealth]
\draw (0,-5) arc (-90:-15:5);
\draw (0,-5) arc (-90:-165:5);
\fill (0,0) circle (2pt);
\draw[thin,->] (0,1) -- (0,-2);
\draw[thin,-<] (0,-2) -- (0,-2.5);
\draw[thin] (0,-2.5) -- (0,-5);
\draw[red] (-3.25,-5) -- (3.25,-5);
\draw[green!75!black,->] (0,0) -- (1.155,-2.5);
\draw[green!75!black] (1.155,-2.5) -- (2.31,-5);
\draw[thin,dashed] (2.31,0) -- (2.31,-5);
\draw[green!75!black,->] (2.31,-5) -- (3.465,-2.5);
\draw[green!75!black] (3.465,-2.5) -- (4.62,0);
\draw[blue,thin] (0,-5) -- (1.90,-4.12);
\draw[blue] (1.70,-4.21) -- (1.765,-4.35) -- (1.965,-4.26);
\draw[orange] (0,-.5) arc (-90:-65.20:.5);
\draw[orange] (.15,-.65) node {\scriptsize $\alpha$};
\draw[orange] (.5,-5) arc (-0:24.80:.5);
\draw[orange] (.65,-4.85) node {\scriptsize $\alpha$};
\draw[orange] (2.31,-4) arc (90:114.80:1);
\draw[orange] (2,-3.75) node {\scriptsize $\alpha$};
\draw[orange] (2.31,-4) arc (90:65.20:1);
\draw[orange] (2.63,-3.75) node {\scriptsize $\alpha$};
\draw[thin,<-] (0,0) -- (-1.6,-1.6);
\draw (-1.765,-1.765) node {\scriptsize $R$};
\draw[thin,->] (-1.93,-1.93) -- (-3.53,-3.53);
\draw[thin,<-] (0,-5.5) -- (1,-5.5);
\draw (1.155,-5.5) node {\scriptsize $\rho$};
\draw[thin,->] (1.31,-5.5) -- (2.31,-5.5);
\end{tikzpicture}
\end{center}
\caption{Light ray deviation from the imperfection of the mirror surface: if the surface is considered locally flat then rays passing through the center of the mirror are not reflected back to that point but are deflected by an angle $\alpha$. In the limit of small angles $\alpha$ one can write $\alpha=\rho/R$, where $\rho$ is the rugosity and $R$ the radius of the sphere.\label{LightRay}}
\end{figure}

Studies \cite{Siegman} of stability of bound modes inside cavities suggest that if the surface of the half-spheres which constitute the mirror were perfect, and in the limit where the paraxial approximation is perfectly valid, the lifetime of the fundamental mode become arbitrarily long. Now, the stability of this (Gau\ss ian) mode can also be characterized in terms of ray optics \cite{Siegman}. In this picture, rays would eternally bounce back and forth between the mirrors. Here, we shall make several simplifying assumptions; we shall assume that rays are straight lines passing though the center of curvature which is situated in the middle of both semi-spherical mirrors. Now, due to the rugosity of the surface of the mirrors, the trajectories would gradually disalign from the radii. In order to simulate this effect, we modelized the rugosity by considering that the surface of the mirrors is not exactly circular but is constituted by juxtaposed planar surfaces. The same kind of approximation was adopted in the discretisation scheme of the quasimodal approach but one should be careful not to confuse the mesh parameter (which is a numerical parameter in the quasimodal approach) with the rugosity parameter considered here, in the context of ray optics, which is given a true physical meaning. Supposing that each time a ray is reflected its angle gets randomly deviated by $\pm 2\rho/R$ 
 (which is the ratio between the rugosity $\rho$ and the radius of curvature $R$ of the mirrors, $R\approx 3 $cm), as illustrated on Fig.~\ref{LightRay}, we find that the impact of the photon on a mirror diffuses according to a random walk process. The randomness can be seen here to reflect the lack of regularity of the rugos surface. After each trip between the mirrors, the photon will come back to its initial position,  modulo a random shift equal to $\pm\left(2\rho/R\right) 2R=\pm4\rho$. As is well known the average distance of diffusion $D$ corresponding to this process obeys $D=\sqrt N 4\rho$ where $N$ is the number of back and forth reflections of the photon. Equalling $D$ with the size of the mirrors (typically 3 cm), we find that $\sqrt N 4\rho=3\cm$. Henceforth, the lifetime, which is equal to $N$ times the duration of a closed trip obeys $\tau=1/\Gamma=\left(3\cm/4\rho\right)^2\times12\cm/c$ where $c$ is the speed of light. This bound delivers a one to one relation between rugosity and lifetime which is
\begin{equation}
\rho={3\cm\sqrt{{12cm\over c \tau}}\over 4}
\end{equation}
The observed lifetime $\tau$ of 130 miliseconds would thus correspond to a rugosity $\rho$ of the order of  $\sqrt 3/4$ micron, more or less 400 nm, in good qualitative agreement with the experimental estimate of the rugosity reported in \cite{Pipeau}, which is 300 nm.

\section{Conclusions and open questions.} \label{sec:Ccl}

In summary we have modelled the losses of a QED cavity thanks to a jump operator which weakly couples the trapped mode inside the cavity to outgoing modes (sect.~\ref{sec:Jump}). Our approach makes it possible to estimate the losses in function of a single parameter which is the tunneling rate between the in and out modes at resonance. Our treatment makes it possible to estimate the spatio-temporal distribution of light throughout time (sect.~\ref{sec:Propag}), and also to characterize the entanglement between the in and out modes (sect.~\ref{sec:Coherent}). As has also been discussed elsewhere, in a first quantization approach \cite{Article}, our approach elucidates the role played by coherent states as classical pointer states (sect.~\ref{sec:Bose}), which opens the way to a classical estimate of the lifetime outlined in sections~\ref{sec:Maxwell} and \ref{sec:rugos}. 

Many ideas developed by us in the present paper are already present in quantum optics textbooks. For instance it is well known that, when an oscillator is linearly coupled to an arbitrary number of oscillators, coherent states factor out during the time evolution. In other words, coherent states are the quasi-classical pointer states of the problem.

A non-standard feature of our treatment is however that we never introduce the concept of a sudden and irreversible quantum jump. This contrasts for instance with a popular approach for deriving the master equation of a lossy QED cavity (\ref{eq:Master}), the so-called Monte-Carlo approach \cite{HarocheRaimond,Dalibard} according to which (here we consider the temperature of the ``environment'', that is to say of the modes outside the cavity, to be equal to zero), after an infinitesimally small time increase $\mathrm{d}t$, either the mode inside the cavity loses one photon (with probability $\Gamma \mathrm{d}t$), or it remains unaffected by the absorption (with probability $1-\Gamma \mathrm{d}t$). This random process can be simulated as a Monte-Carlo process and it leads to the derivation of the master equation  (\ref{eq:Master}) after averaging on the two possible outcomes \cite{Article} (either a detector ``clicks'' and reveals the presence of an outgoing photon or it does not click). Certain features of this model are somewhat puzzling. For instance, in the case where the initial state of the quasi-stationary  mode inside the cavity is a coherent state, it disentangles from the outside world throughout its time evolution, and it is difficult to figure out how the measurement of a photon outside the cavity would affect the state of the mode inside the cavity, even at the mere informational level, because the corresponding modes are totally decorrelated.

In contrast, our approach makes it possible to provide a consistent, unitary description of the evolution of the full system (modes inside AND outside the cavity). In our view, decoherence (which manifests itself through the Lindblad equation (\ref{eq:Master})) is simply seen to be a corollary of entanglement and during the whole process, the state of the full system (cavity+external world outside the boundaries) remains pure and evolves unitarily. Coming back to the famous Schrödinger cat problem, where a cat trapped in a box is prepared into a superposition of a living and of a dead cat, our approach makes it possible to give a complete description of the dead-living cat AND of its direct environment, inside the box. In this view, it is only when we open the box that the superposition breaks down. In a sense, our approach thus makes it possible to provide a  complete description of what happens inside the box. It is only when we open the box that the superposition possibly breaks down.

Besides these foundational issues, the classical techniques that we used for estimating the behaviour of a photon inside the cavity (sects.~\ref{sec:Maxwell} and \ref{sec:rugos}) confirm that classical electromagnetism is sufficient to obtain the main features of the metastable modes inside the cavity. For instance the numerical simulations performed in the quasimodal approach (sect.~\ref{sec:Maxwell}) also predict that the two longest-lived modes inside the cavity are characterized by orthogonal polarizations and possess 9 nodes, and they provide a quantitatively correct estimation of their frequencies as well as of their frequency separation. As can be seen from sect.~\ref{sec:Perfect} however, our estimation of the life times of the trapped modes faces severe problems if we limit us to the quasimode  approach. First, the lifetimes are overestimated and second, the numerical treatment does not converge properly when we increase the mesh parameter. This suggests that in the quasimodal approach either the rugosity or the resistivity is not properly taken into account. A third possibility is that a higher order parametrization of the surface of the cavity mirrors, using curvilinear triangles, will help solve the convergence problem.

Our analysis of sect.~\ref{sec:Loss} shows however that absorption by the mirrors, although it improves the algorithmic convergence of the quasimodal aapproach does not explain the losses because it forces us to overestimate the resistivity of the mirrors. As we show in sect.~\ref{sec:rugos}, we reach a surprisingly good estimate of the life time provided we resort to ray optics in order to properly take account of the rugosity of the mirrors which is a possible source of losses.  We conclude from sect.~\ref{sec:Maxwell} that the main source of losses is the escape of the photons to the free space which surrounds the cavity rather than the absorption by the mirrors, and that the influence of rugosity can be qualitatively explained in terms of a simple model based on ray optics.

As is well-known \cite{Siegman,LaserBeams}, ray optics represents an approximation to the more elaborate description of open resonators in terms of Gau\ss ian beams. This approximation, of course, does not take the radiative leakage phenomena due to the open geometry of the cavity into account. In that respect, the back-of-the-envelope calculation of sect.~\ref{sec:rugos} can be seen as an estimate of the cavity losses caused by the rugosity. It is still an open question to know whether the accuracy of the result is somehow due to numerical luck or highlights the relevance of ray optics presented in \cite{Siegman,LaserBeams}.


\end{document}